\documentclass[10pt]{article}
\usepackage{PRIMEarxiv}

\usepackage[utf8]{inputenc} % allow utf-8 input
\usepackage[T1]{fontenc}    % use 8-bit T1 fonts
\usepackage{hyperref}       % hyperlinks
\usepackage{url}            % simple URL typesetting
\usepackage{booktabs}       % professional-quality tables
\usepackage{amsfonts}       % blackboard math symbols
\usepackage{nicefrac}       % compact symbols for 1/2, etc.
\usepackage{microtype}      % microtypography
\usepackage{lipsum}
\usepackage{fancyhdr}       % header
\usepackage{graphicx}
\usepackage{comment}
\usepackage{physics}
\usepackage{multicol}
\usepackage{multirow}
\usepackage{arydshln}
\usepackage{amssymb}
\usepackage{adjustbox}
\usepackage{physics} % graphics
\graphicspath{{media/}}     % organize your images and other figures under media/ folder
\usepackage{natbib}
\usepackage{color,ulem}  
\definecolor{deepblue}{rgb}{0,0,0.5}  
\definecolor{deepred}{rgb}{0.6,0,0}   
\definecolor{deepgreen}{rgb}{0,0.5,0} 
\definecolor{darkgreen}{rgb}{0,0.6,0} 

\def\ergs{erg s$^{-1}$}
\def\feii{{Fe\sc{ii}}\/}

\def\hb{{\sc{H}}$\beta$\/}
\def\hbbc{{\sc{H}}$\beta_{\rm BC}$}

\def\kms{km s$^{-1}$\/}
\def\l{$\lambda$}

\def\lbol{L$_{\rm bol}$\/}
\def\ledd{L$_{\rm Edd}$}

\def\l5100{L$_{\rm 5100}$}

\def\ciii{{C\sc{iii}]}$\lambda$1909\/}
\def\civ{{C\sc{iv}}$\lambda$1549\/}

\def\redd{R$_{\rm{Edd}}$\/}
\def\rfe{R$_{\rm{FeII}}$\/}

\def\siiv{{Si\sc{iv}}$\lambda$1397\/}

\def\aliii{{Al\sc{iii}}$\lambda$1860\/}
\def\siiii{{Si\sc{iii}]}$\lambda$1892\/}

\title{Preliminary cosmological results using extreme Accretors Quasar formalism
\thanks{\textit{\underline{Citation}}: 
\textbf{Sandoval-Orozco+, Preliminary cosmological results using extreme accretors Quasar formalism. DOI:10.1016/j.dark.2026.102331}} 
}

\author{
  Sandoval-Orozco, R.\\
  Universidad Nacional Aut\'onoma de M\'exico \\
  Instituto de Astronom\'ia \\
  A.P. 70-264, 04510. Ciudad de México, México. \\
  rsandoval@astro.unam.mx\\
   \And
  Negrete C. A. \\
  Universidad Nacional Aut\'onoma de M\'exico \\
  Instituto de Astronom\'ia \\
  A.P. 70-264, 04510. Ciudad de México, México. \\
  SECIHTI Research Fellow \\
   \And 
  Marziani P. \\
  INAF - Osservatorio Astronomico di Padova \\ 
  Osservatorio 5, Padova, IT35122, Italy  \& \\
  Instituto de Astrofísica de Andalucía, \\
  Glorieta de Astronomía, Granada, IAA-CSIC, Spain 
   \And 
   Levi Said J. \\
   Institute of Space Sciences and Astronomy, \\
   University of Malta, Malta, MSD 2080 \\
}

\begin{document}
\maketitle

\begin{abstract}
We revisited the xA Quasar formalism from the cosmological point of view, where a completely cleaned and standardized sample is compiled from different literature references. This allowed us to test three different cosmological models including $w$CDM and $w_0w_a$CDM %\jls{maybe say what these three models are, or something about them} 
and $\Lambda$CDM resulting in a Hubble constant estimation of $H_0 = 69.8 \pm 2.2$~$\mathrm{km\,s^{-1}\,Mpc^{-1}}$ for the compiled sample alone and $H_0=69.0 \pm 0.9$~$\mathrm{km\,s^{-1}\,Mpc^{-1}}$ when combined with Type I Supernovae (SNIa), Cosmic Chronometers and the Cosmic Microwave Background (CMB) distance priors. Using both the $w$CDM and $w_0w_a$CDM a weak Bayesian preference for the dynamical dark energy models over the $\Lambda$CDM model was found. A comparative analysis was performed with other AGN based methods in cosmology like the Reverberation Mapping, the X-Ray and UV non-linear relation and the Angular distance measurements. We conclude that the significant intrinsic dispersion is a key issue present in all samples. Overcoming this dispersion is key to establish xA and other AGN samples as robust and precise cosmological probes.  
\end{abstract}

% keywords can be removed

\section{Introduction}
The cosmological standard model $\Lambda$CDM provides a robust description that successfully accounts for the full range of observational data \citep{di2025cosmoverse}. However, in recent years, several challenges have emerged as the volume and precision of the observational data have improved. Among these challenges lie the determination of key parameters that include: (\textit{i}) The Hubble constant $H_0$ tension \citep{DiValentino:2021izs,CosmoVerseNetwork:2025alb}; (\textit{ii}) the amplitude of matter fluctuations $S_8$ discrepancies \citep{Abdalla:2022yfr}; (\textit{iii}) The determination of the sound horizon at the epoch of reionization \citep{Perivolaropoulos:2024yxv} along with (\textit{iv}) the accurate and precise determination of the $\Omega_m$ and $\Omega_\Lambda$ components. These issues could be indicators of both systematics in the datasets or a hint of new physics. Among the recent results from the DESI survey  \citep{DESI:2024mwx,DESI2025}, there is growing evidence that the cosmological paradigm may require modifications or extensions to serve as a complete cosmological model \citep{di2025cosmoverse}.

In this context, the inclusion of new datasets at high redshifts ($z > 1$), should improve the determination of cosmological parameters and shed light on the nature of dark matter and dark energy. There has been emphasis in resolving the Hubble tension, which requires investigating both potential systematic effects in existing datasets and the possibility of new physics \citep{Perivolaropoulos:2024yxv}. Quasars represent a promising dataset for probing higher cosmological distances, where deviations from the standard cosmological model might become detectable, as suggested by recent analysis using Quasars \citep{trefoloni2024quasars}. Additional methods leveraging quasars as potential standard candles include the ultraviolet-to-X-ray flux relation \citep{Lusso:2020pdb} and reverberation mapping techniques \citep{Panda:2019cvs}. 

We present a compilation of results from multi-year sample analyses of the so called extreme accretor quasars \citep[xA, see Section \ref{sec:xAdescription},][and references therein]{negrete2018highly,Marziani:2014bra,Marziani:2019hzw,Marziani:2013zra}. We argue that these extreme-luminosity objects that irradiate near the Eddington limit\footnote{the Eddington luminosity for those objects is defined as \ledd\ =1.26~$\times 10^{38}$~M/M$_\odot$ \ergs}, when analyzed within the xA formalism, could provide critical insights into the current cosmological paradigm.

While quasars are commonly used alongside other probes in cosmological studies \citep{Dainotti:2024bth}, their dispersion has limited their ability to constrain parameters as effectively as SNIa. Establishing a quasar sample capable of achieving comparable constraints to SNIa from low $z$ but most importantly at $z > 1$, remains an important goal—one that our standardized xA sample aims to advance. A key challenge in previous studies has been sample heterogeneity. In this work, we address this issue by compiling and standardizing scattered samples from the literature to create a well-defined xA cosmological sample.

This paper is organized as follows. Section \ref{sec:xAdescription} contains the description of the potential cosmological application of the Quasars. Section \ref{sec:samples} reports the sample selection and compilation of sources for the different references. Section \ref{sec:cosmological_models} and \ref{sec:methods} present the tested cosmological models and the different cosmological measurements used in this work. In Section \ref{sec:discussion} we present and analyze the results for the xA Quasar sample along with other selected cosmological measurements. Finally, in Section \ref{sec:conclusions} we summarize our results and present perspectives for this work.

\section{Extreme accretors and potential Hubble diagram}
\label{sec:xAdescription}

Quasars are the one of the most luminous sources in the Universe, and therefore their potential for cosmology can be useful to reach up to distances at $z \approx 7$ \citep{Marziani:2013zra}. The problem with using quasars as standard candles is that their range of luminosity is wide. For that reason, \cite{Marziani:2013zra} have proposed the use of a particular kind of quasars, the xA objects, as standard candles. The identification of those objects is based on the Eigenvector 1 (E1) formalism that permits isolating quasars with luminosity near the Eddington limit. The E1 has been built considering four physical properties of the mechanism that fuels the quasar emission, which we call dimensions (4DE1)
%For that reason, we will use the xA approach in which the Eigenvector 1 formalism is used to isolate quasars with luminosity near the Eddington limit. To use this approach, we need to measure four physical properties of the mechanism that fuels the Quasar emission, so called \textit{Eigenvector1}. 

%Eigenvector1 
4DE1 measures the following parameters: 

\begin{itemize}
    \item The velocity dispersion of the broad line emitting region using low-ionization lines, expressed through the Doppler effect on the Full Width at Half Maximum (FWHM) of the broad component (BC) of the \hb\ emission line (\hbbc).
    \item Broad line region ionization and accretion conditions described by the intensity ratio between the \feii\ blend centered at 4570 \AA\ and \hbbc\ known as \rfe. 
    %\item X-ray emission coming from the thermal emission in accretion represented by the index in soft X-rays $\Gamma_\mathrm{soft}$.
    %\item Measurement of non-virial motions in high ionized gas via the amplitude of the \civ\ line blueshift. 
\end{itemize}

%For the distance estimation, we consider the FWHM-H$\beta$ and the $R_{\text{\feii}}$ as tracers
For selecting objects to be used as distance estimators in the low-$z$ regime ($\lesssim$ 0.8), we consider the FWHM(\hbbc) and the \rfe\ as tracers of the physical and dynamical conditions of the BLR \citep{Marziani:2019hzw}. 
%Objects with FWHM $< 4000 $ km/s tend to have strong $R_{\text{\feii}}$ and are  
Objects with FWHM $<$ 4000 \kms\ and strong \rfe\ $>$ 1 are sources that radiate at higher Eddington ratios %$L/L_{\mathrm{Edd}}$.
\redd, defined as the ratio of the bolometric luminosity \lbol, and \ledd. The same selection can be applied to objects up to $z < 1.6$ when the optical range is observed in the infrared. The selection criterion in the UV plane for high-$z$ objects ($\gtrsim$ 2) is based on line ratios of the 1900\AA\ blend. We assume that similar conditions in the accretion regime are present in objects with ratios \ciii/\siiii\ $>$ 1 and \aliii/\siiii\ $>$ 0.5 \citep{Marziani:2014bra}.

The key to using this xA quasars as distance estimators relies on the possibility that we can understand well enough the accretion conditions including that the Broad Line Region is virialized and with no effect from other non-virial motions. These xA objects tend to radiate in a limiting value $L/L_{\mathrm{Edd}} \sim 1$ and their physical conditions along a wide range of $z$ %optical and UV spectra 
%are similar and therefore the radius of emitting regions is proportional as $\propto L^{\frac{1}{2}}$. 
We also require spectral invariance among these objects, as secured by well-defined optical and UV selection criteria, and  the assumption that the region near the black hole
for xA objects require high accretion rates that are only achievable through radiatively inefficient super-Eddington accretion %is virialized
\citep{Marziani:2025ebz} and that there is a spectral invariance in these objects. In summary, we can obtain a luminosity measuring spectroscopic lines as $L \propto$ FWHM$^4$. Details of this derivation are written in \citep{Dultzin:2020lnb}. Essentially, the luminosity can be approximated using: 
\begin{equation}
    L \approx 7.8 \times 10^{44} (\mathrm{FWHM})^4_{1000} \text{ erg s$^{-1}$}.
\end{equation}

The distance modulus computed from the virial luminosity can be written as \citep{Dultzin:2020lnb}: 

\begin{equation}
\label{eq:distance_modulus_measured}
    \mu = 2.5(\log L - \xi )  -2.5\log(f_\lambda \lambda) -2.5\log(4\pi \delta^2_\mathrm{10 pc}) + 5\log(1+z).
\end{equation}
where the constant $-2.5\log(4\pi \delta^2_\mathrm{10 pc}) = -100.19$. The $f_\lambda \lambda$ can be the flux at 5100 \AA\ for the H$\beta$ low-z sample or 1700 \AA\ for the \aliii\ and \siiii\ high-z sample. In this previous expression $\xi$ is the bolometric luminosity correction. Therefore, we can obtain a distance modulus with purely spectroscopic measured quantities.

The criteria to select $R_{\text{\feii }} \geq 1.0$ corresponds to the selection using \aliii$/$\siiii $>0.5$ and \siiii$/$\ciii$>1$ and therefore we can measure objects outside the optical limit for higher redshifts \citep{Dultzin:2020lnb,Marziani:2014bra}. 

\section{Samples description}
%\section{Methods, Observations, Simulations etc.}
\label{sec:samples}

We compile and standardize the following samples. The purpose is to obtain the FWHM of the virialized BCs, \hb\ or \aliii, depending on the $z$ range and the observed spectral interval.

\begin{itemize}
    \item \cite{Marziani:2014bra} objects: The sample consists of 92 objects that span the redshift range between $0.4 \leq z \leq 2.5$. The sample was separated into 
    %two different parts for the analysis, the low redshift catalog $z < 1$ and the high redshift catalog $z > 1$. The low redshift sample FWHM is calculated using the H$\beta$ line while the high redshift is calculated using the ratio \aliii/\siiii/ $\geq$ 0.5.
    three different parts for the analysis, the low redshift catalog $z < 1$ with 58 objects, from the Sloan Digital Sky Survey SDSS data release DR 8, seven intermediate $z$ objects (1.0–2.5) observed in the IR using the VLT/ISAAC instrument, and the high redshift catalog $z > 1$ with 63 objects from the SDSS DR6. %The low and intermediate redshift sample FWHM is calculated using the H$\beta$ line while the high redshift is calculated using the \aliii\ line.
    %\AN{The low redshift sample was selected considering objects with FWHM(\hbbc) $<$ 4000 \kms, and \rfe\ $>$ 1, while the high redshift sample was delimited using the line ratios of the 1900\AA\ blend (Sec. \ref{sec:xAdescription}).}

    For the low redshift sample, the bolometric correction was calculated using \cite{netzer2019bolometric}, which results in BC = 6.3 for this sample. Similar to that, we calculated BC = 6.1 for the high redshift sample. %These objects are referred as \textbf{Marziani \& Sulentic (2014)} in the complete sample.  

    \item \cite{SEAMBH:2016gio} objects: The sample consists of 19 objects in the redshift range spanning between 0.0258 and 0.1364 observed with the Lijiang 2.4 m telescope. All objects in this sample have the FWHM measured using the H$\beta$ line. 

    \item \cite{negrete2018highly} objects: The sample consists of 120 SDSS DR7 objects in the redshift range between 0.08 and 0.71 measuring the \hb line for the FWHM. %The same measurements for the \AN{FWHM(}H$\beta$) for the samples mentioned above are repeated here for a group of selected SDSS DR7 objects. 

    \item \cite{marziani2022intermediate} objects: The sample consists of 28 objects in the low-redshift catalog that measures the FWHM of the H$\beta$ line, between 0.02 and 0.7 in $z$. The high-redshift catalog consists of 20 objects from 1.52 to 3.4 in $z$, measuring the FWHM using the \aliii\ line. The sample was drawn from different sources including the SDSS, 6dF, VLT/ISAAC, \cite{Marziani2003}, and \cite{Sulentic2007}, being careful to avoid duplicating objects from other references. 

    \item \cite{Sniegowska:2020sdh} objects: The sample consists of 13 %objects in the high-redshift catalog that measures the spectral properties using the \aliii\ line,  
    SDSS DR12 high-redshift objects for which their physical properties were determined using the UV emissions of \civ, \siiv, and the 1900 \AA\ blend, in a redshift range between $z \sim 2.1 - 2.3$. The obtained FWHM were using the \aliii\ line.
 
    \item \cite{martinez2018extreme} objects: Consisting of 19 objects observed with the 10m Gran Telescopio de Canarias (GTC), in the redshift range between 2.1 and 2.4 in $z$. Again, the UV lines were used to measure the spectral properties to determine the Super-Eddington luminosity, and the \aliii\ FWHM is measured.

    \item \cite{buendia2025lbt} objects: New sample consisting of 6 objects observed with the near-infrared spectrograph (LUCI) of the Large Binary Telescope (LBT) in the redshift range between 2.3 and 2.4. The FWHM was measured using \hb. %, where \hb\ was observed in the infrared and the 1900\AA\ blend in the optical range. The FWHM was measured using \aliii. 
    %, measuring the FWHM using the \aliii\ line. 

    \item \cite{deconto2023high} objects: Sample containing 3 xA objects in redshift range $z > 2$ measuring the FWHM using the \hb\ line observed in the infrared with the ISAAC spectrograph at VLT. 

    \item \cite{temple2024iii} objects: Sample containing 5 objects obtained using near-infrarred spectra from the Magellan/FIRE covering redshifts from $2.1 < z < 2.3$. The FWHM for this sample is measured directly from the \hb\ line. 

    \item Additional high redshift data points: Three additional objects obtained by analyzing measurements by the James Webb Space Telescope of high-$z$ active galaxies \citep{liu2024fast}, the Gemini Telescope \citep{Shen:2018ojb}, and the DESI Legacy imaging Survey \citep{wang2018discovery}. These objects were re-analyzed in order to obtain new measurements for the sample in a higher redshift span $5.6 \leq  z \leq 7.6$. 

\end{itemize}

The construction of the final sample required a compilation and homogenization process. Data from the aforementioned references were cross-matched to identify common objects. A critical step involved converting all reported measurements into a consistent unit system, specifically expressing the %Full Width at Half Maximum (FWHM) in km/s
FWHM in \kms\ and continuum flux densities in erg/s/cm$^2$, to ensure uniformity across the heterogeneous source material. Following this standardization, the sample was cleansed of potential duplicate entries. For sources with multiple observations at the same redshift, a weighted average was computed to produce a single, representative data point for subsequent analysis Finally, using the different measurements we obtained the distance modulus using equation \eqref{eq:distance_modulus_measured}. 

It is worth noting that this selected sample exhibits several features that merit further discussion. Although the basic tenets of the BLR model that provide the foundation for the cosmological exploitation of xA Quasars, these objects are supported by a large body of the observational evidence and therefore several issues have the potential to introduce systematic errors in the results. An intrinsic effect could be associated with the dependence of the xA SED from black hole mass \citep{franketal02}. Other sources of error could be dependent for every one of the samples used. A major one is the severe bias affecting the highest redshift quasars which could give rise to an "Anscombe-type" statistical bias  \citep{shapovalova12,Anscombe01021973} which may lead to issues in the cosmological results especially in the analysis of the dynamical dark energy scenario. Other effects may increase the dispersion, for example viewing angle effects that are expected to strongly affect the FWHM and the Luminosity \citep{negrete2018highly}. For this reason some of the objects may themselves be subject to a systematic bias (for example, a preferential selection of pole-on sources at higher luminosity).

Nonetheless, the selection of xA sources as cosmological tools have several strong points: first, the self-similarity of the spectrum over a range of 4de1 in luminosity implies some sort of stable, reproducible accretion and Broad Line Region structural configuration, with second order effects due to SED and other parameters. Second: a major result encountered in extensive monitoring campaigns, is the spectroscopic and photometric stability of xA Quasars \citep{shapovalova2012spectral,SEAMBH:2016gio,duetal18}.  Even if the use of a limiting, single Eddington ratio value is an abstraction, the xA sources' Eddington ratio values are distributed around $\sim 1$, with reasonably small dispersion \citep{Marziani:2014bra}. Therefore, the following analysis should be considered still tentative, a step beyond the first approach carried in \citep{Marziani:2014bra}, pending a major assessment of the main sources of systematic and statistical errors that should be carried out in an eventual series of papers.

\section{The cosmological sample}
\label{sec:cosmological_models}
\begin{figure}
    \centering
    \includegraphics[width=0.99\linewidth]{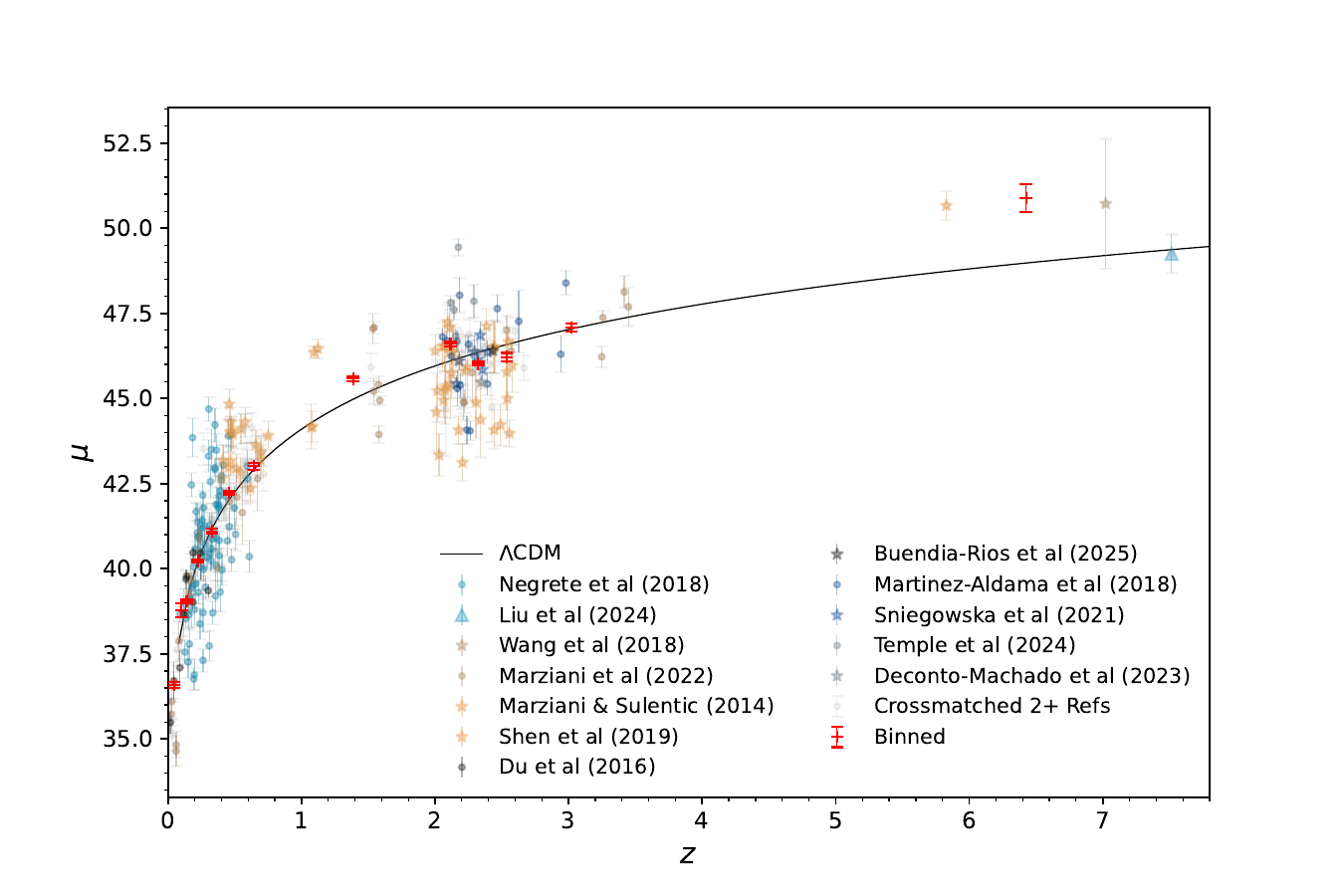}
    \caption{Distance modulus measured with the complete sample with the x-axis as redshift. Different colors represent the samples used for the compilation. The black line represents the $\Lambda$CDM fiducial cosmology with $H_0 = 70$ km/s/Mpc and $\Omega_{m,0} = 0.3$. The red marks show the trend calculated as bins for the complete sample.}
    \label{fig:distance-modulus}
\end{figure}

The final selected sample after the curation comprises 266 objects spanning a redshift range from $z = 0.06$ to $z \sim 7.5$. The distance modulus for this sample is presented in Figure~\ref{fig:distance-modulus}. To facilitate visualization, a binned version of the data was constructed using a modified version of the binning algorithm described in \cite{Brout:2020bbg}. This process reduces scatter and yields a more concise representation of the distance-redshift relation. The binning was performed relative to a fiducial $\Lambda$CDM cosmology with parameters $H_0 = 70~\mathrm{km\,s^{-1}\,Mpc^{-1}}$ and $\Omega_{m,0} = 0.3$.

To determine the cosmological parameters from the xA quasar sample, we perform a Bayesian analysis using a Markov Chain Monte Carlo (MCMC) method. This requires the computation of several key quantities. The cosmological luminosity distance, a fundamental ingredient, is given by:
\begin{equation}
    d_\mathrm{L}(z) = (1 + z) \frac{c}{H_0} \int_0^{z} \frac{dz'}{E(z')},
\end{equation}
where $c$ is the speed of light and $E(z) \equiv H(z)/H_0$ is the normalized Hubble parameter.

We first consider the $\Lambda$CDM model, for which $E(z)$ takes the form:
\begin{equation}
    E^2(z) = \Omega_{m,0} (1 + z)^3 + \Omega_{r,0}(1+z)^4 + \Omega_{\Lambda,0},
\end{equation}
where $\Omega_{m,0}$, $\Omega_{r,0}$, and $\Omega_{\Lambda,0}$ are the present-day density parameters for matter, radiation, and dark energy, respectively. The parameter vector for this model is $\Theta = (H_0, \Omega_{m,0})$.

We also extend our analysis to dynamical dark energy models. The $w$CDM parametrization is characterized by:
\begin{equation}
    E^2(z) = \Omega_{m,0} (1 + z)^3 + \Omega_{r,0}(1+z)^4 + \Omega_{\Lambda,0} (1 + z)^{3(1+w_0)},
\end{equation}
where $w_0$ is the constant dark energy equation of state parameter. The free parameters for this model are $\Theta = (H_0, \Omega_{m,0}, w_0)$.

Finally, we consider the Chevallier-Polarski-Linder (CPL) parametrization \citep{Chevallier:2000qy}, %Linder:2002et}:
\begin{align}
    E^2(z) & = \Omega_{m,0} (1 + z)^3 + \Omega_{r,0}(1+z)^4 \\ & + \Omega_{\Lambda,0} (1 + z)^{3(1+w_0 + w_a)}\exp\qty(\frac{-3w_az}{1+z}),
\end{align}
which introduces a redshift-dependent equation of state $w(z) = w_0 + w_a z / (1+z)$. The parameter vector is $\Theta = (H_0, \Omega_{m,0}, w_0, w_a)$. This model will be referred to as $w_0w_a$CDM.

For all models, the distance modulus is calculated as:
\begin{equation}
    \mu(z, \Theta) = 5 \log_{10}\left( \frac{d_\mathrm{L}(z, \Theta)}{1 \mathrm{Mpc}} \right) + 25,
    \label{eq:distance_modulus}
\end{equation}
where the luminosity distance $d_\mathrm{L}$ is expressed in units of Mpc.

The log-likelihood function for the xA quasar sample is then defined as:
\begin{equation}
    \ln \mathcal{L}_\mathrm{xA} = - \frac{1}{2} \sum_{i}^{N=266} \left[ \frac{ \left( \mu(z_i, \Theta) - \mu_{\mathrm{obs}, i} \right)^2 }{ \sigma_{\mu, i}^2 } + \ln \left( 2\pi \sigma_{\mu, i}^2 \right) \right],
\end{equation}
where $\mu_{\mathrm{obs}, i}$ and $\sigma_{\mu, i}$ are the observed distance modulus and its uncertainty for the $i$-th quasar, respectively.

Previous attempts of using the xA formalism to obtain a suitable cosmological sample are present in literature. In \cite{Marziani:2014bra} the sample was used to constraint the cosmological parameters as $\Omega_m = 0.19^{+0.16}_{-0.08}$ which has a larger uncertainties but consistent with later studies using a compilation of samples reported in \cite{Czerny_2021} were the authors obtained $\Omega_m = 0.290^{+0.048}_{-0.043}$ assuming a $H_0$ fixed at the Planck measurements \citep{Planck:2018vyg}.  

For our analysis, the MCMC sampling is performed using the \textsc{nautilus} algorithm \citep{nautilus}, and the resulting posterior distributions are analyzed and visualized with the \textsc{getdist} package \citep{Lewis:2019xzd}.

Within the $\Lambda$CDM framework, the full xA sample constrains the cosmological parameters to $H_0 = 68.8 \pm 2.2~\mathrm{km\,s^{-1}\,Mpc^{-1}}$ and $\Omega_{m,0} = 0.264^{+0.054}_{-0.046}$. The binned sample yields consistent values of $H_0 = 69.3^{+2.1}_{-2.2}~\mathrm{km\,s^{-1}\,Mpc^{-1}}$ and $\Omega_{m,0} = 0.251^{+0.049}_{-0.047}$ at the $2\sigma$ confidence level, demonstrating agreement between the two data treatments. Consequently, all subsequent cosmological tests reported herein utilize only the full sample to avoid redundancy. For the $w$CDM model, the xA Quasar sample provides the constraints $H_0 = 76.2 \pm 4.7~\mathrm{km\,s^{-1}\,Mpc^{-1}}$, $\Omega_{m,0} = 0.288 \pm 0.043$, and $w_0 = -2.19 \pm 0.63$. Finally, for the $w_0w_a$CDM model, the constraints are $H_0 = 75.4 \pm 4.8~\mathrm{km\,s^{-1}\,Mpc^{-1}}$, $\Omega_{m,0} = 0.294^{+0.046}_{-0.041}$, $w_0 = -1.98^{+0.71}_{-0.78}$, and $w_a = -1.5^{+2.8}_{-1.6}$. It is important to remember that those determinations are in the context of the aforementioned issues with the cosmological distances determination. 

These results for the xA Quasar sample will be analyzed in detail in the context of those derived from other cosmological probes, with a particular focus on comparisons against constraints obtained from alternative Quasar-based approaches.

%\section{Binning and Clipping sample}

%\RS{
%For the representation of the dataset we will use a modified version of \cite{Brout:2020bbg} binning method in order to obtain a cleaner and smaller sample for visualization. In this case the fiducial model will be the $\Lambda$CDM model with $H_0 = 70$ km/s/Mpc and $\Omega_{m,0} = 0.3$.
%}

%\begin{figure}
%    \centering
%    \includegraphics[width=0.99\linewidth]{binned.pdf}
%    \caption{Binned sample in red points. The full sample is plotted in light gray while the $\Lambda$CDM with $H_0 = 70$ km/s/Mpc and $\Omega_{m,0} = 0.3$ is represented in the black line. X-axis represents redshift while y-axis represents the distance modulus. }
%    \label{fig:binned}
%\end{figure}

\section{Sample comparison with other cosmological measurements}
\label{sec:methods}

We will compare the cosmological constraints obtained from the xA Quasar sample with those derived from other Quasar or AGN-based approaches. To provide a robust context for this comparison, the quasar datasets will be complemented with standard cosmological probes. This combined analysis will facilitate a comprehensive assessment of the performance and consistency of state-of-the-art quasar methods for measuring cosmological distances. For that we will include several samples:

\begin{itemize}
    \item \textbf{Reverberation mapping time-delay}. We utilize the $R-L$ relation derived from reverberation mapping. The dataset, comprising 157 measurements of the flux at 5100\,\AA\ and the time delay $\tau$ from the H$\beta$ line, is adopted from \cite{Cao:2025ldl}. The analysis employs the following log-likelihood function:
    \begin{equation}
    \ln \mathcal{L}_\mathrm{nRL} = - \frac{1}{2} \sum_{i}^{N} \left[ \left( \frac{\tau_{\text{H}\beta,i} - \beta_{RL} - \gamma_{RL} \log L_{5100,i}}{\sigma_{\text{tot},i}} \right)^2 + \ln(2 \pi \sigma_{\text{tot},i}^2) \right],
    \end{equation}
    where $L_{5100} = 4\pi d_\mathrm{L}^2(z,\Theta) F_{5100}$ denotes the monochromatic luminosity, $N = 157$ is the sample size, and $\sigma_{\text{tot},i}$ represents the total uncertainty for the $i$-th measurement. The total uncertainty is defined as:
    \[
    \sigma_{\text{tot},i}^2 = \sigma_{\text{int}}^2 + \sigma_{\log \tau_{\text{H}\beta},i}^2 + \gamma_{RL}^2 \sigma^2_{\log F_{5100,i}},
    \]
    with $\sigma_{\text{int}}$ treated as a free parameter to account for intrinsic scatter. Asymmetric uncertainties in the original data are handled conservatively by adopting the larger value from the upper and lower error bars for each data point, following similar methodology of \cite{Cao:2025ldl}. The prior probability distributions for the nuisance parameters associated with this sample are provided in Table~\ref{tab:priors}. This dataset will be referred to as the \textit{nRL} sample.
    
    \item \textbf{non-linear UV-X sample}. We employ the $L_\mathrm{X} - L_\mathrm{UV}$ relation for a sample of 2421 quasars spanning the redshift range $0.009 \leq z \leq 7.52$ \citep{Lusso:2020pdb,Lusso:2025bhy}. Assuming the relation to be redshift-independent, it is expressed in terms of observed fluxes as:
    \begin{align}
    \log F_\mathrm{X}(z,\Theta) &= \beta_{\mathrm{uvx}} + \gamma_\mathrm{uvx}(\log F_\mathrm{UV} + 27.5) \nonumber \\
    & - 2(1 - \gamma_\mathrm{uvx}) (\log d_\mathrm{L}(z,\Theta)-28.5)
    + (\gamma_\mathrm{uvx} -1)\log(4\pi),
    \end{align}
    The corresponding log-likelihood function is given by \citep{Lusso:2025bhy}:
    \begin{small}
            \begin{align}
        \ln \mathcal{L}_\mathrm{nUVX} = -\frac{1}{2} \sum_i^N \left[\frac{\left(\log F_{\mathrm{X},i}^{\text{obs}} - \log F_{\mathrm{X}}(z_i,\Theta)\right)^2}{\left(\mathrm{d}\log F_{\mathrm{X},i}^{\text{obs}} \right)^2+ e^{2 \ln \delta}} + \ln\left[\left(\mathrm{d}\log F_{\mathrm{X},i}^{\text{obs}} \right)^2+ e^{2 \ln \delta}\right]\right],
    \end{align}
    \end{small}

    where the parameter $\delta$ characterizes the intrinsic dispersion of the relation. The prior distributions for all parameters are listed in Table~\ref{tab:priors}. This sample will be referred to as the \textit{nUVX QSO} sample in the text and figures. This formulation implies that the nUVX QSO sample can only be used in combination with the SNIa sample, as it requires an external calibration for the absolute distance scale.

    %For the present analysis, we adopt a modified approach in which the parameters $\beta_{\mathrm{uvx}}$ and $\gamma_\mathrm{uvx}$ are fixed to the values $\beta_\mathrm{uvx} = -31.475  \pm 0.008$ and $\gamma_\mathrm{uvx} = 0.591 \pm 0.011$, as determined by \cite{Benetti:2025ljc}. Under this assumption, the luminosity distance $d_\mathrm{L}$ (and consequently the distance modulus $\mu$) can be directly constrained. The likelihood function thus simplifies to:
    %\begin{equation}
    %\ln \mathcal{L}_\mathrm{nUVX} = -\frac{1}{2} \sum_i^N \left[\frac{\left(\mu_{i}^\mathrm{obs} -\mu_i(z_i,\Theta,k)\right)^2}{\sigma^2_\mathrm{tot}} + \ln \left(2\pi \sigma^2_\mathrm{tot}\right)\right],
    %\end{equation}
    %where the total variance is $\sigma^2_\mathrm{tot}=\sigma_{\mu_i^{\text{obs}}}^2 + \delta^2$. The nuisance parameter $k$ is introduced to account for zero-point calibration uncertainties within the quasar sample. 

    \item \textbf{Angular distance measurements}. We incorporate ultra-compact radio sources observed via Very Long Baseline Interferometry (VLBI). The sample consists of 120 sources spanning a redshift range of $0.474 < z < 2.743$, as presented in \cite{Cao:2017ivt}. These objects serve as standard rulers, where the angular size $\theta(z)$ is related to the cosmological angular diameter distance $d_\mathrm{A}(z)$ and the intrinsic physical length $\ell_\mathrm{m}$ by the relation:
    \begin{equation}
        \theta(z) = \frac{\ell_\mathrm{m}}{d_\mathrm{A}(z)}.
    \end{equation}
    We adopt a Gaussian prior on the intrinsic length scale of $\ell_\mathrm{m} = 11.03 \pm 0.25$ pc to constrain the cosmological model, thereby introducing $\ell_\mathrm{m}$ as a nuisance parameter. The methodology for determining this intrinsic size is detailed in the original literature \citep{Cao:2017ivt}.

    The corresponding log-likelihood function used to constrain the cosmological parameters is given by:
    \begin{equation}
        \ln \mathcal{L}_\mathrm{AngularQso} = -\frac{1}{2} \sum_{i}^{120} \left[ \frac{ \left( \theta(z_i, \Theta; \ell_\mathrm{m}) - \theta_{\mathrm{obs}, i} \right)^2 }{ \sigma_{\text{tot}, i}^2 } + \ln \left( 2 \pi \sigma_{\text{tot}, i}^2 \right) \right].
    \end{equation}
    Here, $\Theta$ represents the set of cosmological parameters. The total uncertainty for the $i$-th measurement, $\sigma_{\text{tot}, i}$, incorporates a 10\% contribution ($0.1\theta_{\mathrm{obs}, i}$) to account for the empirical dispersion inherent in the dataset and the angular size measurements. The prior probability distributions for the parameters associated with this sample are provided in Table~\ref{tab:priors}. This dataset will be referred to as the \textit{Angular QSO} sample.
\end{itemize}

As several of the individual quasar samples lack the statistical robustness to constrain the full set of cosmological parameters independently, we incorporate three complementary cosmological datasets. The inclusion of these external probes is essential to ensure robust parameter estimation and to facilitate convergence in MCMC sampling for the Quasar-based methods. Those samples are:
\begin{itemize}
    \item \textbf{Pantheon+ SH0ES SNIa sample}. The Pantheon+ sample comprises 1701 confirmed Type Ia supernova (SNIa) light curves, corresponding to 1550 distinct objects, spanning a redshift range of $0.001 < z < 2.3$ \citep{Scolnic:2021amr}. This dataset provides the necessary observables to compute the distance modulus, defined as $\mu = m - M$, where $M$ is the absolute magnitude.

    A key feature of the Pantheon+ sample is its calibration using Cepheid variables in nearby host galaxies. The residuals, $\Delta \mu$, between the observed and model distance moduli are therefore calculated differently for calibrated and uncalibrated supernovae:
    \begin{equation}
        \Delta \mu(z_i,\Theta) = \begin{cases}
        \mu_\text{obs}(z_i) - \mu_i^\text{Cepheids} & i \in \text{Cepheid host} \\
        \mu_\text{obs}(z_i) - \mu(z_i,\Theta) &i \notin \text{Cepheid host}. 
        \end{cases}
    \end{equation}
    where $\mu_i^\text{Cepheids}$ is the distance modulus anchored by Cepheid measurements. This treatment follows the standard methodology outlined in the foundational Pantheon+ papers \citep{Scolnic:2021amr}.
     The log-likelihood function for the cosmological parameters $\Theta$ is constructed from these residuals and the full covariance matrix $C_{ij}$, which encapsulates both statistical and systematic uncertainties:
    \begin{equation}
        \ln \mathcal{L}_\mathrm{SN} = -\frac{1}{2} \left[ \Delta \vec{\mu}(\Theta)^T \, \mathbf{C}^{-1} \, \Delta \vec{\mu}(\Theta) \right],
    \end{equation}
    where $\mathbf{C}^{-1}$ is the inverse of the covariance matrix \citep{Brout:2020bbg}. The prior distributions for any associated nuisance parameters are provided in Table~\ref{tab:priors}. This dataset will be referred to as the \textit{SNIa} sample in the subsequent analysis.

    \item \textbf{Hubble parameter measurements}. We considered the observational measurements of the Hubble parameter, $H(z)$, obtained from the Cosmic Chronometers (CC) technique. The sample consists of 31 data points spanning the redshift range $0.3 < z < 2.0$ \citep{Moresco:2016mzx}. This method constrains the Hubble parameter by measuring the differential age evolution of passively evolving galaxies within the same cluster, utilizing small differences in their redshifts. The stellar population synthesis models used to estimate the ages of these galaxies are detailed in the literature \citep{Moresco2024}. The dataset will be analyzed through the likelihood: 
    \begin{equation}
        \ln \mathcal{L}_\mathrm{CC} = -\frac{1}{2} \left[ \Delta H(z_i,\Theta)^T \, \mathbf{C}^{-1} \, \Delta H(z_i,\Theta) \right],
    \end{equation}
    where $\Delta H(z_i,\Theta) = H_\mathrm{obs}(z_i) - H(z_i, \Theta)$ is the residual between the observed and the model-predicted Hubble parameter at redshift $z_i$, and $\mathbf{C}^{-1}$ is the inverse of the covariance matrix, which accounts for measurement errors and systematic uncertainties \citep{Moresco:2020fbm}. This dataset will be referred to as the \textit{CC} sample in the forthcoming analysis.

    \item \textbf{CMB distance priors}. To probe the constraints on the early universe, we incorporate the compressed Cosmic Microwave Background (CMB) likelihood from the final Planck data release \citep{Chen:2018dbv}. These distance priors provide an efficient means to test cosmological models against early-universe physics without requiring the full computation of the perturbative equations.
    The method utilizes a vector of observables, $\mathbf{V}(\Theta) = (R, \ell_\mathrm{A}, \Omega_b h^2)$, which consists of the shift parameter ($R$), the acoustic scale ($\ell_\mathrm{A}$), and the physical baryon density ($\Omega_b h^2$) where $h = H_0/100 \mathrm{\,km\,s^{-1}\,Mpc^{-1}}$. These quantities are defined as follows:
    \begin{align}
        R &= \sqrt{\Omega_m} \frac{H_0}{c} d_\mathrm{A}(z_*), \\
        \ell_\mathrm{A} &= (1 + z_*) \frac{\pi d_\mathrm{A}(z_*)}{r_s(z_*)},
    \end{align}
    where $c$ is the speed of light, $d_\mathrm{A}(z_*) = d_\mathrm{L}(z_*)/(1+z_*)^2$ is the angular diameter distance at the redshift of decoupling $z_*$, and $r_s(z_*)$ is the comoving sound horizon at that redshift.

    The redshift of decoupling, $z_*$, is approximated by the fitting formula:
    \begin{equation}
        z_* = 1048 \left[ 1 + 0.00124(\Omega_b h^2)^{-0.738} \right] \left[1 + g_1 (\Omega_m h^2)^{g_2} \right],
    \end{equation}
    where the factors $g_1$ and $g_2$ are given by:
    \begin{align}
        g_1 &= \frac{0.0738(\Omega_b h^2)^{-0.238}}{1 + 39.5(\Omega_b h^2)^{0.763}}, \\
        g_2 &= \frac{0.560}{1 + 21.1(\Omega_b h^2)^{1.81}}.
    \end{align}
    The constraints from this dataset are derived through the log-likelihood:
    \begin{equation}
        \ln \mathcal{L}_\mathrm{CMB} = - \frac{1}{2} \Delta \mathbf{V}^T \, \mathbf{C}^{-1} \, \Delta \mathbf{V},
    \end{equation}
    where $\Delta \mathbf{V} = \mathbf{V}_\mathrm{obs} - \mathbf{V}(\Theta)$ is the residual vector and $\mathbf{C}^{-1}$ is the inverse of the covariance matrix for the Planck 2018 data release \citep{Chen:2018dbv}. This dataset will be referred to as the \textit{CMB Light} sample.

    \end{itemize}

\begin{table}
    \centering
    \begin{tabular}{ccc}
    \hline
    \hline
        Case & Parameter & Prior \\
        \hline
        \multirow{4}{*}{ All } & $H_0$ [$\mathrm{km\,s^{-1}\,Mpc^{-1}}$] & Uniform$(50,100)$ \\
           & $\Omega_{m,0}$ & Uniform$(0,1)$ \\
           & $w_0$ & Uniform$(-3,0)$ \\ 
           & $w_a$ & Uniform$(-3,3)$ \\ 
        \hline 
        \hline
        \multirow{3}{*}{nRL} & $\beta_{RL}$ & Uniform$(0,10)$ \\
            & $\gamma_{RL}$ & Uniform$(0.5)$ \\
            & $\sigma_{\text{int}}$ & Uniform$(0,5)$\\
        \hline
        \multirow{2}{*}{nUVX Qso} & $\delta$ & Uniform$(0,2)$ \\
         & $k$ & Uniform$(-1,1)$ \\
         %& $\gamma_\text{nuvx}$ & Uniform$(0.4,0.8)$ \\
         %& $\beta_\text{uvx}$ & Uniform$(-40,-20)$ 
         %& $K$ & Uniform$(-1,1)$ \\
        \hline
        Angular Qso & $\ell_\mathrm{m}$ [pc] & Gaussian$(11.03,0.25)$ \\
        \hline
        SNIa & $M$ & Uniform$(-20,-17)$ \\ 
        \hline 
        CMB \textit{Light} & $\Omega_bh^2$ & Uniform$(0.01,0.04)$ \\
        \hline
    \end{tabular}
    \caption{Priors for the different nuisance and cosmological parameters for the different analyzed samples.}
    \label{tab:priors}
\end{table}

\begin{figure}
    \centering
    \includegraphics[width=0.8\linewidth]{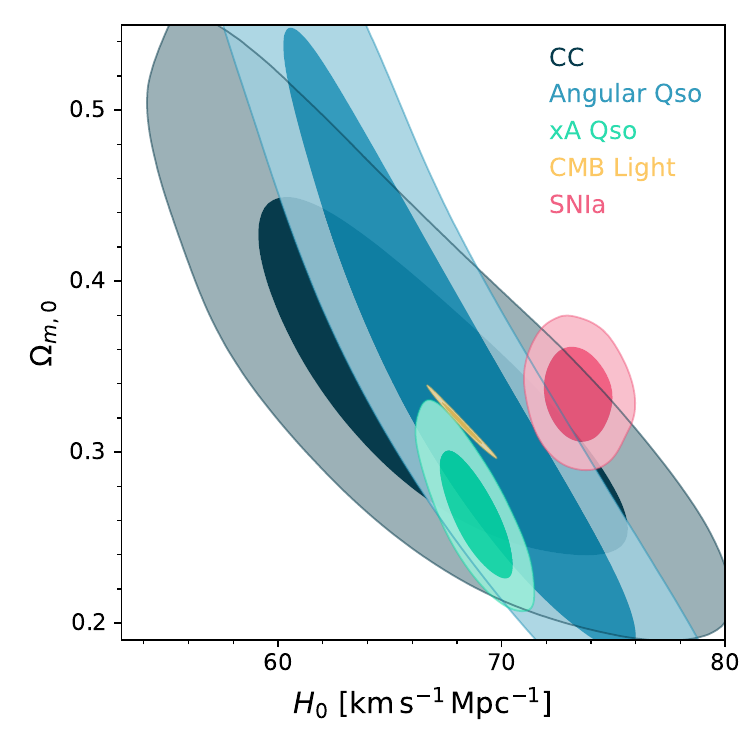}
    \caption{Parameter space of $H_0$ and $\Omega_{m,0}$ for the $\Lambda$CDM model tested for with xA Qso sample showing the results using CMB light, the Pantheon+ SNIa, the Cosmic Chronometers analysis and the Angular Quasar sample. Those are the samples that are capable of constraining the cosmological parameters individually.}
    \label{fig:base_lcdm}
\end{figure}

\begin{figure}
    \centering
    \includegraphics[width=0.8\linewidth]{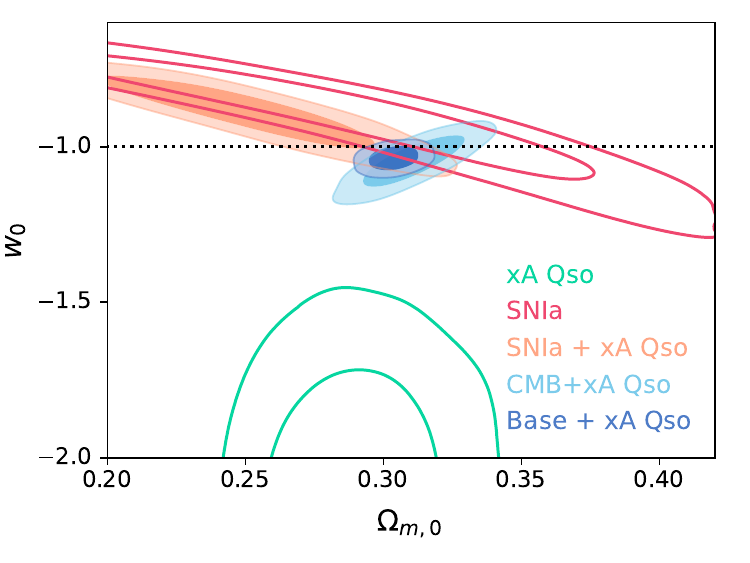}
    \caption{Bi-dimensional contour plots for the $w$CDM model in the $\Omega_{m,0} - w_0$ plane at 1 and 2$\sigma$ confidence levels obtained for each dataset combination marked in the label. The marked dashed line represents where $w_0 = 1$ or the $\Lambda$CDM confirmation. }
    \label{fig:xA_wcdm}
\end{figure}

\begin{table}
    \centering 
    \begin{adjustbox}{width=\columnwidth,center}
    \begin{tabular}{ccccccccc}
    \hline
    \textbf{Model} &  & Dataset & $H_0$  & $\Omega_{m,0}$ & $w_0$ & $w_a$ & $\Omega_bh^2$ & $\ln \mathcal{B}_{ij}$ \\
    & & & [$\mathrm{km\,s^{-1}\,Mpc^{-1}}$]   & \\
    \hline
    $\boldsymbol{\Lambda}$\textbf{CDM}&  & & & & -1 (fixed) & 0 (fixed) & \\
    \hline
    & \multirow{ 4}{*}{\rotatebox[origin=c]{90}{\textbf{xA Qso}}} & Individual & $68.8 \pm 2.2$ & $0.264^{+0.054}_{-0.046}$ & -- & --& -- & -- \\
     & & + SNIa & $69.9\pm 1.3$ & $0.290\pm 0.027 $ & -- & -- & -- & -- \\
     & & + CMB \textit{Light} & $68.1 \pm 1.0$ & $0.319 \pm 0.015$& -- & -- & $0.02232 \pm 0.00027$ & -- \\
     & & + Base & $69.0 \pm 0.9$ & $0.306 \pm 0.012$ & -- & -- & $0.02249 \pm 0.00025$ & -- \\[1mm]
    \hdashline
     & \multirow{ 4}{*}{\rotatebox[origin=c]{90}{\textbf{\scriptsize{Angular Qso}}}} & Individual & $67.0 \pm 10.0$ & $0.38^{+0.29}_{-0.23}$ & -- & -- & -- & -- \\[0.1mm]
     & & + SNIa & $72.3 \pm 1.8$ & $0.318 \pm 0.033$ & -- & -- & -- & -- \\ 
     & & + CMB \textit{Light} & $68.2 \pm 1.2$ & $0.317 \pm 0.017$ & -- & -- & $0.02236 \pm 0.00029$ & -- \\ 
     & & + Base & $69.34 \pm 0.97$ & $0.302 \pm 0.013$ & -- & -- & $0.02256 \pm 0.00027$ & -- \\ 
     \hdashline 
     &\multirow{ 2}{*}{\rotatebox[origin=c]{90}{\textbf{nUVX}}} &%Individual & $75 \pm 20$ & $0.929^{+0.076}_{-0.14}$ & -- & -- & -- & --  \\
     + SNIa & $73.3 \pm 2.0$ & $0.352 \pm 0.037$ & -- & -- & -- & -- \\
     %& & + CMB \textit{Light} & $67.9 \pm 1.2$ & $0.321 \pm 0.017$ & -- & -- & $0.02230 \pm 0.00030$ & -- \\
     & & + Base & $69.2 \pm 1.0$ & $0.305 \pm 0.013$ & -- & -- & $0.02253 \pm 0.00026$ & -- \\[1mm]
    \hdashline
     &\multirow{ 4}{*}{\rotatebox[origin=c]{90}{\textbf{nRL}}} & Individual & $75 \pm 20$ & $0.53 \pm 0.49$ & -- & -- & -- & -- \\
     & & + SNIa & $73.4 \pm 2.0$ & $0.334 \pm 0.035$ & -- & -- & -- & -- \\
     & & + CMB \textit{Light} & $68.2 \pm 1.2$ & $0.317 \pm 0.017$ & -- & -- & $0.02235 \pm 0.00029$ & -- \\
     & & + Base & $69.34 \pm 0.96$ & $0.302 \pm 0.013$ & --& -- & $0.02256 \pm 0.00026$  & -- \\
    \hline
%%%%%%%%%%%%%%%%%%%%%%%%
    $\boldsymbol{w}$\textbf{CDM}&  & & & &  & 0 (fixed) & \\
    \hline
     & \multirow{ 4}{*}{\rotatebox[origin=c]{90}{\textbf{xA Qso}}} & Individual & $76.2 \pm 4.7$ & $0.288 \pm 0.043$ & $-2.19 \pm 0.63$ & -- & -- & $-7.58$ \\
     & & + SNIa & $69.8 \pm 1.3$ & $0.249 \pm 0.067 $ & $-0.89 \pm 0.16$ & -- & -- & $+1.64$ \\
     & & + CMB \textit{Light} & $69.1 \pm 2.7$ & $0.310 \pm 0.024$ & $-1.05 \pm 0.11$ & -- & $0.02227 \pm 0.00030$ & $+2.77$ \\
     & & + Base & $69.6 \pm 1.1$ & $0.304 \pm 0.012$ & $-1.036 \pm 0.050$ & -- & $0.02240 \pm 0.00028$ & $+2.83$ \\
     \hdashline 
    & \multirow{ 4}{*}{\rotatebox[origin=c]{90}{\textbf{\scriptsize{Angular Qso}}}} & Individual & $73 \pm 20$ & $0.34^{+0.33}_{-0.26}$ & $-1.6 \pm 1.3$ & -- & -- & $+0.08$ \\
     & & + SNIa & $72.2 \pm 1.8$ & $0.24 \pm 0.12$ & $-0.83\pm 0.22 $ & -- & -- & $+1.13$ \\ 
     & & + CMB \textit{Light} & $74^{+20}_{-10}$ & $0.28 \pm 0.12$ & $-1.16^{+0.46}_{-0.53}$  & -- & $0.02235 \pm 0.00029$ & $+1.50$\\ 
     & & + Base & $69.9 \pm 1.3$ & $0.299 \pm 0.014$ &$-1.032 \pm 0.052$ & -- & $0.02249 \pm 0.00029$ & $+3.17$ \\ 
     \hdashline 
     &\multirow{ 2}{*}{\rotatebox[origin=c]{90}{\textbf{nUVX}}} %& Individual & $75 \pm 20$ & $0.89^{+0.12}_{-0.19}$ & $-1.6^{+1.6}_{-1.4}$ & -- & -- & $-0.32$\\
     & + SNIa & $73.6 \pm 2.1$ & $0.428^{+0.069}_{-0.075}$ & $-1.31^{+0.30}_{-0.33}$ & -- & -- & $-0.11$ \\
     %& & + CMB \textit{Light} &  $51.3^{+3.2}_{-1.4}$ & $0.563^{+0.038}_{-0.055}$ & $-0.381^{+0.060}_{-0.12}$ & -- & $0.02236 \pm 0.00029$ & $+7.28$ \\
     & & + Base & $69.6 \pm 1.3$ & $0.302 \pm 0.013$ & $-1.028\pm 0.050$ & -- & $0.02247 \pm 0.00028$ & $+3.24$ \\[1mm]
    \hdashline
     &\multirow{ 4}{*}{\rotatebox[origin=c]{90}{\textbf{nRL}}} & Individual & $75 \pm 20$ & $0.55 \pm 0.49$ & $-1.4^{+1.3}_{-1.5}$ & -- & -- & $+0.05$ \\
     & & + SNIa & $73.3 \pm 2.0$ & $0.28 \pm 0.14$ & $-0.90 \pm 0.30$ & -- & -- & $+1.72$ \\
     & & + CMB \textit{Light} & $72^{+30}_{-20}$ & $0.32^{+0.23}_{-0.18}$ & $-1.08 \pm 0.78$ & -- & $0.02236 \pm 0.00029$ & $+0.73$ \\
     & & + Base & $69.9 \pm 1.3$ & $0.299 \pm 0.013$ & $-1.030 \pm 0.051$ & -- & $0.02250 \pm 0.00028$ & $+3.13$ \\
    \hline
%%%%%%%%%%%%%%%%%%%%%%%%%%%%
    $\boldsymbol{w_0w_a}$\textbf{CDM}&  & & & & & & \\
    \hline
    & \multirow{ 4}{*}{\rotatebox[origin=c]{90}{\textbf{xA Qso}}} & Individual & $75.4 \pm 4.8$ & $0.294^{+0.046}_{-0.041}$ & $-1.98^{+0.71}_{-0.78}$ & $-1.5^{+2.8}_{-1.6}$ & -- & $-8.07$ \\
    & & + SNIa & $69.5 \pm 1.3$ & $0.305^{+0.078}_{-0.10}$ & $-0.81 \pm 0.18$ & $-1.5^{+2.1}_{-1.5}$ & -- & $+1.36$ \\
    & & + CMB \textit{Light} & $71.7 \pm 3.1$ & $0.289 \pm 0.24$ & $-1.52^{+0.24}_{-0.20}$ & $1.37^{+0.37}_{-0.54}$ & $0.02230 \pm 0.00032$ & $-0.54$ \\
    & & + Base & $69.8 \pm 1.2$ & $0.303 \pm 0.012$ & $-0.91 \pm 0.14$ & $-0.60 \pm 0.65$ & $0.02232 \pm 0.00029$ & $+3.04$ \\
    \hdashline
    & \multirow{ 4}{*}{\rotatebox[origin=c]{90}{\textbf{\scriptsize{Angular Qso}}}} & Individual & $74.0 \pm 20.0$ & $0.34^{+0.32}_{-0.27}$ & $-1.6^{+1.4}_{-1.3}$ &  $-0.3^{+3.0}_{-2.6}$ & -- & $+0.14$ \\
     & & + SNIa & $72.1 \pm 1.8$ & $0.26^{+0.15}_{-0.21}$ & $-0.82^{+0.21}_{-0.23}$ & $-0.5^{+1.5}_{-2.0}$ & -- & $+2.01$ \\ 
     & & + CMB \textit{Light} & $74 \pm 20$ & $0.28 \pm 0.15$ & $-1.1 \pm 1.1$ & $-0.5 \pm 2.3$ & $0.2235 \pm 0.00030$ & $+1.76$ \\ 
     & & + Base & $70.8 \pm 1.4$ & $0.293 \pm 0.014$ & $-0.76 \pm 0.15$ & $-1.29 \pm 0.78$ & $0.02240 \pm 0.00028$ & $-1.39$ \\ 
     \hdashline 
     &\multirow{ 2}{*}{\rotatebox[origin=c]{90}{\textbf{nUVX}}} %& Individual & $74 \pm 20$ & $0.23^{+0.32}_{-0.23}$ & $-0.11^{+0.12}_{-0.20}$ & $2.76^{+0.26}_{-0.44}$ & -- & $-10.11$ \\
     &  + SNIa & $73.4 \pm 2.1$ & $0.423 \pm 0.077$ & $-1.07 \pm 0.30$ & $-1.7^{+1.7}_{-1.4}$ & -- & $-6.57$ \\
     %& & + CMB \textit{Light} & $56.8^{+4.4}_{-3.6}$ & $0.458^{+0.057}_{-0.069}$ & $-0.65^{+0.27}_{-0.29}$ & $0.14^{+0.50}_{-0.67}$ & $0.02234 \pm 0.00028$ & $+5.67$\\
     & & + Base & $70.5 \pm 1.3$ & $0.296 \pm 0.013$ & $-0.81 \pm 0.14$ & $-1.06 \pm 0.71$ & $0.02240 \pm 0.00029$ & $+1.01$ \\[1mm]
    \hdashline
     &\multirow{ 4}{*}{\rotatebox[origin=c]{90}{\textbf{nRL}}} & Individual & $75 \pm 20$ & $0.54 \pm 0.48$ & $-1.4 \pm 1.4$ & $0.0 \pm 2.9$ & -- & $+0.04$ \\
     & & + SNIa & $73.3^{+2.1}_{-2.0}$ & $0.31^{+0.16}_{-0.23}$ & $-0.90^{+0.27}_{-0.29}$ & $-0.6^{+1.7}_{-2.2}$ &  -- & $+2.36$ \\
     & & + CMB \textit{Light} & $76 \pm 20$ & $0.28^{+0.17}_{-0.13}$ & $-1.01^{+0.93}_{-0.97}$ & $-0.96^{+2.4}_{-2.1}$ & $0.02234 \pm 0.00031$ & $+1.17$ \\
     & & + Base & $70.9 \pm 1.4$ & $0.292 \pm 0.014$ & $-0.75 \pm 0.16$ & $-1.35^{+0.73}_{-0.80}$ & $0.02242 \pm 0.00029$ & $-1.80$ \\ 
    \hline
    \end{tabular}
    \end{adjustbox}
    \caption{Results of the cosmological parameters constraints for the studied samples reported at 95\% confidence level. \textit{Base} refers to the Combination of CMB distance priors, SNIa and Cosmic Chronometers. The evidences are calculated as $\ln \mathcal{B}_{ij} = \ln \mathcal{Z}_{\Lambda\mathrm{CDM}} - \ln\mathcal{Z}_{\mathrm{Model}}$ where Model is both $w$CDM or $w_0 w_a$CDM for the same dataset combinations. A positive value shows preference for the $\Lambda$CDM model while a negative value shows preference for the dynamical Dark Energy. The nuissance parameters for every dataset are calculated but not shown for simplicity reasons. }
    \label{tab:results}
\end{table}

\begin{figure}
    \centering
    \includegraphics[width=0.8\linewidth]{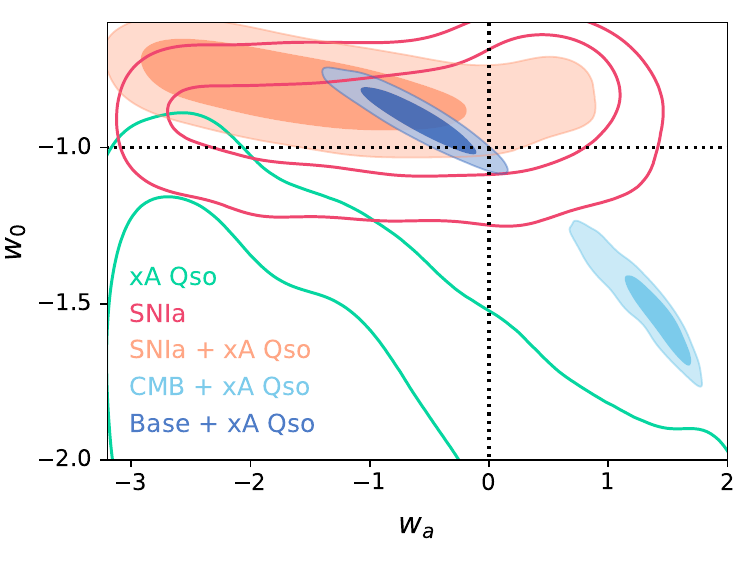}
    \caption{Bi-dimensional contour plots for the $w_0w_a$CDM model tested in the $w_0 - w_a$ plane at 1 and 2$\sigma$ confidence levels obtained from each dataset combination marked in the label. The marked dashed lines represents $w_0 = 1$ and $w_a = 0$ or the $\Lambda$CDM confirmation. }
    \label{fig:xA_cpl}
\end{figure}

\section{Discussion}
\label{sec:discussion}

\begin{figure}
    \centering
    \includegraphics[width=0.99\linewidth]{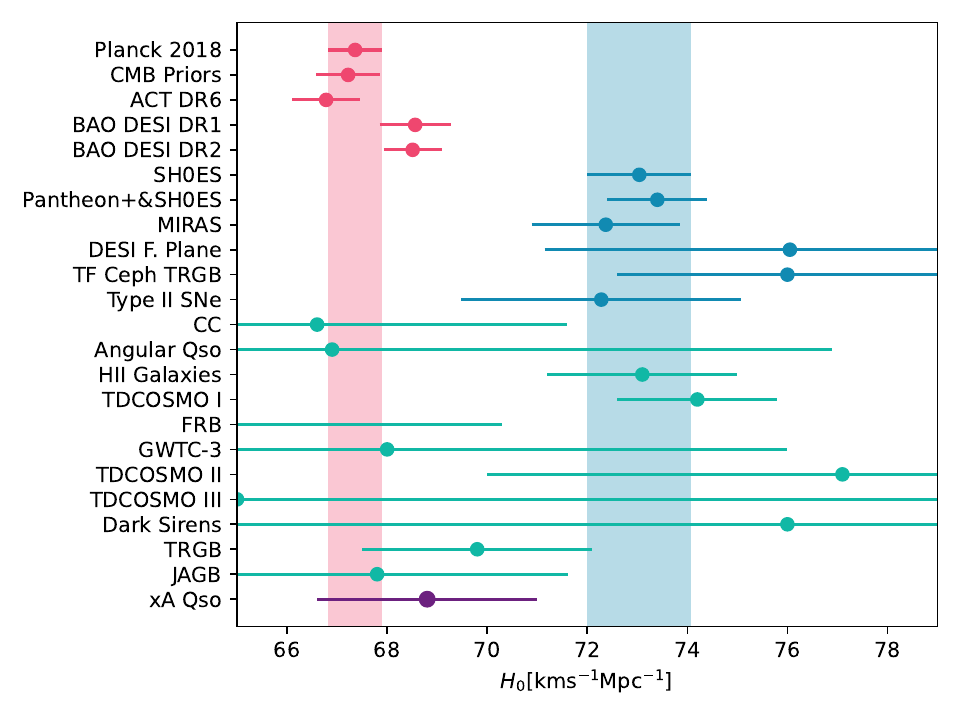}
    \caption{Whisker plot for the different $H_0$ estimations in the literature. Blue color represents measurements that depend on Cepheids distance ladder, red color the ones that depend on the early universe, and green ones the late-time determinations not dependent on distance ladder assumptions. We considered the measurements Planck 2018 \citep{Planck:2018vyg}, CMB distance priors \citep{Chen:2018dbv}, Atacama Cosmology Telescope (ACT) DR6 \citep{ACT:2025fju}, and both BAO DESI \citep{DESI:2024mwx,DESI2025} for the early inferences of the Hubble constant. For the late-time distance ladder dependent measurements we considered the SH0ES calibration \citep{Riess:2021jrx}, the Pantheon+ and SH0ES determination \citep{Brout:2020bbg}, the MIRA Distances \citep{Huang:2023frr}, the DESI fundamental plane determination \citep{Scolnic:2024hbh}, the Tully-Fisher relation calibration with Cepheids and Tip of the Red Giant Branch (TRGB) \citep{Scolnic:2024oth} and the determination using TypeII SNe \citep{deJaeger:2023vkm}. For the late-time non dependent on distance ladder we considered the Cosmic Chronometers \citep{Moresco:2020fbm}, the Angular Quasars used in this work \citep{Cao:2017ivt}, the HII regions estimation \citep{Chavez:2024twa}, different TDCOSMO approaches \citep{Millon:2019slk,TDCOSMO:2023hni,TDCOSMO:2024rwr}, the dark siren measurements \citep{Alfradique:2023giv}, TRGB and J-Region Asymptotic Giant Branch (JAGB) calibration \citep{Freedman:2024eph}, the Gravitational Wave Transient Catalog determination \citep{LIGOScientific:2021aug} and the FRB estimations \citep{Wu:2021jyk}. Dark colored regions are the measurements from Planck and SH0ES.  }
    \label{fig:whisker}
\end{figure}

The xA Quasar sample constraints the Hubble constant to $H_0 = 68.8 \pm 2.2~\mathrm{km\,s^{-1}\,Mpc^{-1}}$ within the $\Lambda$CDM model, which is an intermediate value between the estimates of Planck \citep{Planck:2018vyg} and SH0ES \citep{Riess:2025chq}. A critical distinction of the xA Quasar sample is the absence of a zero-point calibration analogous to the Cepheid-supernova distance ladder. Consequently the constraints on $H_0$, derived solely from this sample are inherently less precise and exhibit larger uncertainties. A comparison of the cosmological constraints from the xA quasar sample with those derived from the Pantheon+, Cosmic Chronometers, CMB distance priors, and Angular QSO samples is presented in Figure~\ref{fig:base_lcdm}. The xA quasar sample demonstrates consistency with these established cosmological probes across the parameter space $H_0-\Omega_{m,0}$. The context among other determinations of the Hubble constant $H_0$ is detailed in Figure~\ref{fig:whisker} where the determination is consistent with the early universe determinations and the methods that do not depend on distance ladder assumptions \citep{Perivolaropoulos:2024yxv}.  

For the $w$CDM model, the constraint on the Hubble constant shifts to $H_0 = 76.2 \pm 4.7~\mathrm{km\,s^{-1}\,Mpc^{-1}}$, showing consistency with the SH0ES measurement, though with a significantly larger uncertainty. Conversely, the matter density parameter is constrained to $\Omega_{m,0} = 0.288 \pm 0.043$, which is more precise than the $\Lambda$CDM result. The equation of state parameter is found to be $w_0 = -2.19 \pm 0.63$. While this constraint is relatively weak due to the substantial uncertainty, it is noteworthy that the Bayesian evidence indicates a statistical preference for a dynamical dark energy model over a cosmological constant ($w = -1$). The constraints from the $w_0w_a$CDM model yield $H_0 = 75.4 \pm 4.8~\mathrm{km\,s^{-1}\,Mpc^{-1}}$ which again is consistent with the SH0ES estimation but with a large error margin. The matter density estimation $\Omega_{m,0} = 0.294^{+0.046}_{-0.041}$ is fully consistent with the simpler $\Lambda$CDM model. Both dark energy parameters are $w_0 = -1.98^{+0.71}_{-0.78}$ and $w_a = -1.5^{+2.8}_{-1.6}$.  Although the central values deviate from the $\Lambda$CDM expectation ($w_0=-1, w_a=0$), the large uncertainties prevent any decisive conclusion. This trend, where the data hints at dynamics but lacks the precision to rule out a cosmological constant, is consistent with recent findings from the DESI BAO measurements \citep{DESI:2024mwx, DESI2025}, which also suggest a slight preference for evolving dark energy. However, it is important to consider that these results may also reflect an intrinsic dispersion within the Quasar sample itself (like the aforementioned lack of zero point calibration), which could limit the precision achievable for constraining dynamical dark energy parameters. This potential source of uncertainty warrants further investigation in future studies. 

To contextualize the potential intrinsic dispersion of the xA sample, we compare its constraints with those from other Quasar-based cosmological approaches. Given the distinct methodologies and selection functions detailed in Section~\ref{sec:methods}, these samples are analyzed separately rather than combined. The cosmological parameter constraints for all models and datasets are summarized in Table~\ref{tab:results}; for simplicity, we report only the cosmological parameters, although the full posteriors including nuisance parameters were sampled.

The combination of the Pantheon+, CMB distance priors, and Cosmic Chronometers (CC) datasets is denoted as the \textit{Base} sample. To quantitatively assess the preference for different dark energy models, we compute the Bayesian evidence, $\ln \mathcal{Z}$, for each model. The relative evidence is then expressed via the Bayes factor, defined as $\ln B_{ij} = \ln \mathcal{Z}_{\Lambda\mathrm{CDM}} - \ln\mathcal{Z}_{\mathrm{Model}}$.

Our comparative analysis of Quasar cosmological probes yields several key insights. A primary finding is that the xA Quasar sample is the only approach capable of constraining the full parameter set for all three cosmological models independently. This sample consistently yields a matter density parameter of $\Omega_{m,0} \sim 0.3$ across all models. When combined with CMB distance priors, all four quasar samples converge to similar cosmological parameter values, suggesting a degree of underlying consistency. Notably, the preference for dynamical dark energy over $\Lambda$CDM using xA Quasar with the rest of the selected datasets vanishes, which instead favor the standard model. The combinations for the xA sample are plotted in bi-dimensional parameter space $\Omega_{m,0} - w_0$ for the $w$CDM model and $w_0 - w_a$ for the $w_0w_a$CDM model in the Figures~\ref{fig:xA_wcdm} and~\ref{fig:xA_cpl}. 

In contrast, the nUVX quasar sample produces also constraints that show a preference for dynamical dark energy, with $w < -1$ in the $w$CDM model and $w_a \neq 0$ in the $w_0w_a$CDM parametrization.  Its combination with SNIa strongly favors a high Hubble constant $H_0 = 73.3 \pm 2.0$, and it exhibits a significant internal tension when combined with the CMB distance priors, resulting in an anomalous, poor fit ($H_0 = 51.3^{+3.2}_{-1.4}$, $\Omega_m = 0.563^{+0.038}_{-0.055}$). This is probably a result of the lack of the BAO inclusion for the CMB data \citep{Chen:2018dbv} with dark energy parametrizations. This is repeated under both extended dark energy models. However the lack of constraints is resolved by the full Base combination that strongly favors $\Lambda$CDM similar to the xA Quasar sample. The combination with Pantheon+, Cosmic Chronometers and CMB distance priors bring the parameters back to consensus values ($H_0 = 69.2 \pm 1.0$, $\Omega_m = 0.305 \pm 0.013$). Crucially, the bayesian evidence for the Base combination favors $\Lambda$CDM over the $w$CDM model ($\ln \mathcal{B}_{ij} = +3.24$) and provides positive evidence over the $w_0w_a$CDM model ($\ln \mathcal{B}_{ij} = +1.01$). All the evidence could indicate that higher-redshift data favor alternatives to $\Lambda$CDM, though not conclusive. For most dataset combinations, the parameter $w_a$ remains poorly constrained, and its inclusion is not statistically justified. More details about the nUVX sample are in \cite{Benetti:2025ljc}.

Both the Reverberation Mapping Quasars and Angular Quasar samples adopt higher $H_0$ values in combination with the SNIa sample and lower $H_0$ values with CMB distance priors. This is due to the lack of constraint for the cosmological parameters using the sample individually. In the case of the Angular Quasars the evidence using the Base dataset is positive toward the $\Lambda$CDM model in the case of $w$CDM. This is not the case for the $w_0w_a$CDM model where the evidence is negative ($\ln \mathcal{B}_{ij} = -1.39$) suggesting a slight preference for the dynamical dark energy. The nRL sample where the combination with the Base datasets results in a weakly preference for the CPL parametrization ($\ln \mathcal{B}_{ij} = -1.80$). This, depict the large error margin for the free parameter $w_a = -1.35^{+0.73}_{-0.80}$. Again, similar to the other Quasar analysis, this could be a hint of dispersion and not inherently because a deviation from the standard cosmological model is needed. 

The overall picture when combining Quasar data with established probes as SNIa, Cosmic Chronometers, and CMB reinforces the standard cosmological model except for the Angular. These robust combinations constrain the dark energy equation of state to be consistent with a cosmological constant ($w = -1$) within $2\sigma$ confidence levels. Furthermore, the Bayesian evidence prefers $\Lambda$CDM over extended dark energy models for most of the combinations. The large uncertainties on $w_0$ and $w_a$ in analyses that suggest dynamics often indicate a lack of constraining power rather than a definitive detection of deviation. This, coupled with the fact that some Quasars-driven datasets prefer a higher $H_0$ while others like xA are consistent with lower values, suggests that the Hubble tension may be more influenced by data-specific systematics than by the cosmological model. %As expected, combinations including the CMB yield the tightest constraints on the $\Lambda$CDM parameters. Ultimately, while individual quasar samples hint at intriguing possibilities, they currently lack the statistical robustness to challenge the prevailing cosmological model.

%The estimation for $\Omega_{m,0} = 0.264^{+0.054}_{-0.046}$ for the xA sample is consistent in $2\sigma$ with the CMB and the SNIa constraints. Yet, the xA calibration contains a bigger error than the other two samples for both the cosmological parameters. 

\section{Conclusions}
\label{sec:conclusions}

A common characteristic across all quasar samples is their significant intrinsic dispersion. This substantial scatter likely contributes to the inability of these datasets to provide tight constraints on more complex models, such as $w$CDM and $w_0w_a$CDM. These findings underscore the importance of developing physically better-understood quasar samples, as a reduction in intrinsic dispersion is crucial for leveraging them as precise cosmological probes. This is crucial specially for the xA sample as a bigger sample with less dispersion is crucial to improve the accuracy in the cosmological parameter determination. 

These results demonstrate that the cosmological constraints, particularly the Hubble constant $H_0$, are highly sensitive to the specific AGN sample and standardization technique employed. The heterogeneity in outcomes across different Quasar approaches indicates that they are not yet a homogeneous cosmological tool. This variability likely stems more from the evolving physical understanding of quasars and their systematic uncertainties than from an inherent issue with the $\Lambda$CDM model.

Our analysis finds weak evidence for dynamical dark energy ($w \neq -1$) when considering the xA Quasar sample in isolation, a finding consistent with other recent studies \citep{Chaudhary:2025uzr}. However, this preference vanishes in the most robust analyses that combine the xA sample with Pantheon+, CMB distance priors, and Cosmic Chronometers. These combined datasets yield strong Bayesian evidence in favor of the standard $\Lambda$CDM model. The apparent deviations observed in some quasar-only analyses are therefore more plausibly attributed to systematic uncertainties in the standardization methods than to new physics.

Key systematic challenges include the significant intrinsic dispersion within the samples and, for the xA Quasars, a lack of data in the critical redshift ranges $1 < z < 2$ and $z > 3$. This gaps in the dataset could be an issue to test the different models for the dynamical dark energy. Future efforts to develop xA quasars into precision cosmological probes must prioritize understanding and reducing this dispersion and unevenly distribution, as well as filling these observational gaps with other observational techniques. Systematic studies aiming to mass determination and accretion rate are also a priority. A full physical and dynamical model for the core regions of black holes are a crucial necessity and those are not available for the moment, despite the fact recent GRAVITY observations suggest a possible confirmation for mutually independent virialized  and wind dynamical subsystems  that are assumptions for the xA approach \citep{dayem2025spatially}. Additionally the determination of the luminosity is highly dependent on the $L_0$ choice. %This choice is assumed based on the assumption that $L/L_{\mathrm{Edd}} \approx 1$ but there could be changes in the SED of the Quasar. For example, the photoionization could not contribute to the assumed spectral invariance.  

The terms entering the definition of $L_0$ (which is assumed constant) are mainly related to the shape of the ionising SED of the xA sources (see \citealt{Marziani:2014bra} for a thorough discussion). They should not be directly dependent on $H_0$.  The Eddington ratio -- also entering in the definition of $L_0$ --  could be considered   a free parameter with a prior distribution  (Gaussian, skewed Gaussian etc), with modest dispersion (Figure 5 of \citet{Marziani:2014bra}. The terms   associated with the ionizing SED. In the adopted value of L0 they are assumed as appropriate for high Eddington ratio and fairly high luminosity, a convenient  first approximation. Major  changes in L0 induced by the SED are not expected: otherwise, in a photoionization context, spectral invariance would not be preserved: spectra satisfying the condition defining xA sources are found from very low redshift up to the highest redshift sampled by JWST observations (see the section on the cosmological sample for a brief discussion of possible systematic effects). At any rate, L0 as a  constant should be independently estimated via the Hubble diagram fit. Therefore all the parameters used to calculated the relation $L \sim 7.8 \times 10^{44} (\mathrm{FWHM})^4_{1000} \mathrm{erg s}^{-1}$ are slightly dependent on physical assumptions including the black hole mass that are going to be studied in a forthcoming paper.

Our next approaches will include attempts to fill this observational gaps, studies on the calibration of the virial luminosity and possible effects on the host contribution of the xA samples. 

All the aforementioned comments regarding the xA sample should be considered before using the sample, specially in comparison with other well-established ones as the nUVX, where the study of the possible systematics have been done with higher precision. 

Nevertheless we recommend that this sample should be considered as a potential cosmological dataset considering the current challenges on the systematics discused previously.

A comprehensive investigation into extended cosmological models, including non-flat universes and modified gravity scenarios with the xA sample, also will be presented in a forthcoming paper.

\section*{Acknowledgements}

PM acknowledges financial support from the Spanish MCIU through project PID2022-140871NB-C21 by “ERDF A way of making Europe”, and from the Severo Ochoa grant CEX2021-515001131-S funded by MCIN/AEI/10.13039/501100011033. JLS would also like to acknowledge funding from “Xjenza Malta” as part of the “Technology Development Programme” DTP-2024-014 (CosmicLearning) Project.CAN
acknowledges the support from grants SECIHTI CBF2023-2024-141852 (2026) 102331 DGAPA PAPIIT IA-104325, IN-112226, IN-113026, IN-113726. 

We would also like to thank Feije Wang for kindly providing to the authors the necessary spectra for one of the high-redshift objects. His help allowed us to unlock a crucial redshift as part of the xA Quasar sample.

%%%%%%%%%%%%%%%%%%%%%%%%%%%%%%%%%%%%%%%%%%%%%%%%%%
\section*{Data Availability}

Data Availability Statements provide a standardised format for readers to understand the availability of data underlying the research results described in the article.

%Bibliography
\bibliographystyle{apalike}  
\bibliography{references}

\end{document}